\documentclass[a4paper,11pt]{article}
\usepackage{jheppub} % for details on the use of the package, please see the JINST-author-manual
\usepackage[title]{appendix}
\usepackage{cancel}
\usepackage{xcolor}
\usepackage{tensor}
\title{\boldmath Hydrodynamical transports in generic AdS Gauss-Bonnet-scalar Gravity }

\author[a]{Chenwei Tong,}
\author[b]{Rohit Mishra,}
\author[c]{Yanqi Wang}
\author[c,d]{and Song He\footnote{Corresponding author.}}
\affiliation[a]{Center for Theoretical Physics and College of Physics, Jilin University,\\
 Changchun 130012, China}
\affiliation[b]{Department of Physics,\\
R.K.Mission Vivekananda Centenary College,\\
Rahara, Kolkata, India}
\affiliation[c]{Institute of Fundamental Physics and Quantum Technology,\\
\& School of Physical Science and Technology,\\
Ningbo University, Ningbo, Zhejiang 315211, China}
\affiliation[d]{Max Planck Institute for Gravitational Physics (Albert Einstein Institute),\\ 
Am M\"uhlenberg 1, 14476 Golm, Germany}

\emailAdd{tongcw24@mails.jlu.edu.cn}
\emailAdd{mrohit996@gmail.com}
\emailAdd{wangyanqi0@gmail.com}
\emailAdd{hesong@nbu.edu.cn}

\abstract{The experimentally observed temperature-dependent shear and bulk viscosities of the quark-gluon plasma (QGP), along with its apparent violation of the Kovtun-Son-Starinets (KSS) bound $\eta/s=1/(4\pi)$, necessitate a holographic description that incorporates higher-derivative corrections. We propose a five-dimensional Einstein-Scalar-Maxwell-Gauss-Bonnet model in which a scalar-Gauss-Bonnet coupling $H(\phi)$ encodes leading curvature corrections. Although no closed-form black hole solution is available, we employ an entropy-production analysis at the event horizon to derive exact analytic formulas for the shear viscosity $\eta$ and bulk viscosity $\zeta$. These expressions exhibit apparent deviation from the KSS bound and nontrivial temperature dependence. We then perform an independent computation via the retarded Green function (Kubo) method, finding perfect agreement for $\eta$ and isolating a single constant in $\zeta$ that requires numerical determination. Our dual derivation underscores the pivotal role of higher-derivative terms in realistic QGP modeling and demonstrates the efficacy of nonanalytic holographic backgrounds in capturing the dynamics of strongly coupled fluids.}

\begin{document}
\maketitle
%\flushbottom

%{
%\section{Introduction}
%It is no news that fluid dynamics can be studied by holography.
%}
\section{Introduction and summary}
Relativistic fluid dynamics is a prominent area of research in theoretical physics, with applications spanning astrophysics \cite{Alford:2017rxf, Most:2022yhe, Shternin:2008es, Gursoy:2017wzz}, cosmology \cite{Iosifidis:2024ksa, Murphy:1977ub, Bemfica:2022dnk}, and nuclear physics \cite{Busza:2018rrf}. In astrophysics, it plays a crucial role in understanding phenomena such as neutron star formation and evolution, where gravitational effects are so strong that non-relativistic physics cannot be applied. In cosmology, relativistic fluid models describe the expansion and cooling of the early universe. In high-energy nuclear physics, the quark-gluon plasma (QGP) produced in heavy-ion collisions is characterized as a nearly perfect fluid \cite{Heinz:2013th}.  

The properties of relativistic fluids can be studied both phenomenologically and theoretically. Phenomenological approaches rely on heavy-ion collision data analyzed using Bayesian methods \cite{Nijs:2020ors, Nijs:2020roc, Nijs:2023yab, Giacalone:2023cet, JETSCAPE:2020mzn, Apostolidis:2025gnn}. Models such as the M$\ddot{\text{u}}$ller-Israel-Stewart formalism \cite{Denicol:2014vaa} and the Trento model \cite{Moreland:2014oya} are used to match experimental data. One key discovery is the temperature dependence of the ratios $\eta/s$ (shear viscosity to entropy density) and $\zeta/s$ (bulk viscosity to entropy density), despite the absence of magnetic or electric fields in these models.  

Theoretically, relativistic fluids can be studied holographically, where the fluid resides on the boundary of a gravitational theory. In this framework, various aspects of relativistic fluids can be studied, including dispersion relations \cite{Denicol:2008rj, Baggioli:2024zfq} and transport coefficients. In this work, we focus on the latter, which are computed using diverse methods, notably fluid/gravity correspondence and linear response theory. The fluid/gravity correspondence, pioneered in~\cite{Bhattacharyya:2007vjd, Bhattacharyya:2008xc}, provides an analytical tool for deriving the stress-energy tensor  
\begin{align}
    T_{\mu\nu} = \epsilon u_\mu u_\nu + p P_{\mu\nu} - 2\eta \sigma_{\mu\nu}-\zeta \partial u P_{\mu\nu}
\end{align}
directly,  where $\epsilon$ is the energy density, $p$ is the pressure, $u_\mu$ is the fluid velocity, $P_{\mu\nu} = \eta_{\mu\nu} + u_\mu u_\nu$ is the projection operator, $\sigma_{\mu\nu}$ represents the shear tensor as $\sigma_{\mu\nu}=\tensor{P}{^\rho_\mu}\tensor{P}{^\sigma_\nu}\nabla_{(\rho}u_{\sigma)}-\frac{1}{3}P_{\mu\nu}\nabla u$. Recent applications of fluid/gravity correspondence primarily focus on Chamblin-Reall models \cite{Wu:2016pmx, Wu:2023nlw} (involving a single scalar field), as additional fields or charges introduce differential equations that yield non-analytic solutions \cite{Torabian:2009qk}. Linear response theory offers a broader approach, computing retarded Green's functions via conserved fluxes \cite{Policastro:2001yc, Son:2007vk,Li:2014dsa,Villarreal:2025ydo} as $\operatorname{Im} G_R=-2\mathcal{F}$, where $\mathcal{F}$ relates to perturbations of the metric, or calculated by boundary stress-energy tensors as \cite{Grozdanov:2016tdf, Wang:2024bli} 
\begin{align}
    \delta T^{\mu\nu}=-\frac{1}{2} G^{\mu\nu,\lambda\sigma}_{R}\delta g_{\lambda\sigma}.
\end{align}
Within this framework, shear viscosity $\eta$ is determined from off-diagonal metric perturbations, while bulk viscosity $\zeta$ is extracted from diagonal perturbations \cite{Natsuume:2014sfa}. Both methods require solving perturbation equations of motion, which can be highly complex.  

A renowned result in holographic fluid dynamics is the KSS bound \cite{Kovtun:2003wp}, which states $\eta / s = 1/(4\pi)$. While this bound holds in many holographic models, it is not universal; experimental results show temperature-dependent ratios. Holography offers several mechanisms to violate the KSS bound, including magnetic fields \cite{Wang:2024bli}, anisotropic backgrounds \cite{Erdmenger:2010xm}, scalar-tensor terms (see \cite{Feng:2015oea, Bravo-Gaete:2021hlc, Bravo-Gaete:2022lno, Bravo-Gaete:2023iry} for example, where \cite{Feng:2015oea, Bravo-Gaete:2021hlc} are about Horndeski theory, \cite{Bravo-Gaete:2022lno} discussed the Degenerate
Higher-Order-Scalar-Tensor (DHOST) theory), and higher-derivative corrections \cite{Cremonini:2011iq, Apostolidis:2025gnn, Brigante:2007nu,Buchel:2024umq}. Motivated by experimental approaches that involve neither extra fields nor anisotropy, we study models with temperature-dependent transport coefficients. Higher-derivative terms thus become the natural choice. Existing literature often focuses on models with analytic metric solutions \cite{Ge:2009eh,Ge:2008ni} or perturbative Gauss-Bonnet treatments \cite{Bu:2015bwa, Buchel:2023fst, Buchel:2024umq}. We investigate non-perturbative higher-derivative corrections and derive analytic expressions for shear and bulk viscosities. A major challenge is the absence of analytic metric solutions in such models. Fortunately, an alternative method based on entropy variation \cite{Fairoos:2018pee} enables calculation of first-order transport coefficients ($\eta$ and $\zeta$) without an exact solution. In this paper, we compute the shear and bulk viscosities for the five-dimensional  Einstein-Scalar-Maxwell-Gauss-Bonnet (ESMGB) theory \cite{Boulware:1986dr, Caceres:2023gfa}. We demonstrate that transport coefficients can be calculated using the aforementioned methods, yielding results consistent with the Kubo formula.  

This paper is structured as follows: In Section \ref{sec:intro}, we introduce the ESMGB theory and present its equations of motion. By substituting these into the action, we express the on-shell action as a total derivative, as discussed in \cite{Buchel:2023fst}. Section \ref{sec:thermo} calculates thermodynamic quantities, including Hawking temperature, entropy, energy density, and pressure. The latter two are determined holographically from the diagonal components of the boundary stress-energy tensor. In Section \ref{sec:transport}, we compute shear and bulk viscosities using the Kubo formula for later comparison with the entropy-production method. Section \ref{sec:entropy} introduces the entropy-production method and derives viscosities for simpler models (e.g., Einstein-Maxwell-Dilaton and Gauss-Bonnet theories). Section \ref{sec:GB} presents viscosity calculations for the ESMGB model. Finally, Section \ref{sec:conclusion} summarizes our findings and outlines future research directions.

\section{Holographic model}
\label{sec:intro}
In this section, we introduce our holographic model and present the corresponding equations of motion. The action for the five-dimensional Einstein-Maxwell-Dilaton theory with a Gauss-Bonnet coupling term is given by
\begin{equation}
\label{EMDGBAction}
    S = \frac{1}{2 \kappa_N^2} \int d^5x \sqrt{-g} \left[ \mathcal{R} - \frac{1}{2} \nabla_\mu \phi \nabla^\mu \phi - \frac{Z(\phi)}{4} F_{\mu\nu} F^{\mu\nu} - V(\phi)+\alpha H(\phi)\mathcal{R}_{GB}^{2} \right],
\end{equation}
where $\mathcal{R}^2_{\text{GB}} = \mathcal{R}^2 - 4 \mathcal{R}_{\mu\nu} \mathcal{R}^{\mu\nu} + \mathcal{R}_{\mu\nu\alpha\beta} \mathcal{R}^{\mu\nu\alpha\beta}$ is the Gauss-Bonnet invariant, and $\mathcal{R}$ is the Ricci scalar associated with the metric $g_{\mu\nu}$. The scalar field is denoted by $\phi = \phi(r)$. The potential $V(\phi)$ and the coupling functions $Z(\phi)$ and $H(\phi)$ depend solely on $\phi$, while $\alpha$ is the Gauss-Bonnet coupling constant. A key distinction of our model compared to previous studies \cite{Apostolidis:2025gnn} is the inclusion of a scalar field coupling in the Gauss-Bonnet term. The equations of motion are as follows:
\begin{equation}\label{phieq}
    \nabla_{\mu} \nabla^{\mu} \phi-\frac{\partial_{\phi} Z}{4} F_{\mu \nu} F^{\mu \nu}-\partial_{\phi} V+\alpha \partial_{\phi} H \mathcal{R}^2_{\text{GB}}=0,
\end{equation}
\begin{equation}\label{eleceq}
    \nabla^{\nu}\left(\sqrt{-g} Z F_{\nu \mu}\right)=0,
\end{equation}
\begin{align}
    &\mathcal{R}_{\mu\nu}-\frac{1}{2}\mathcal{R}g_{\mu\nu}+\alpha \left(H(\phi)H_{\mu\nu}+4P_{\mu\alpha\nu\beta}\nabla^{\alpha}\nabla^\beta H(\phi)\right)\cr
    &=\frac{1}{2}\partial_{\mu}\phi\partial_{\nu}\phi+\frac{Z}{2}F_{\mu\rho}F_{\nu}^{\rho}+\frac{1}{2}\left(-\frac{1}{2}\nabla_{\mu}\phi\nabla^{\mu}\phi-\frac{Z}{4}F_{\mu\nu}F^{\mu\nu}-V\right)g_{\mu\nu},\label{EOME}
\end{align} 
where $H_{\mu\nu}$ and $P_{\mu\nu\rho\sigma}$ are defined by
\begin{align}
    H_{\mu\nu}=&-\frac{1}{2}g_{\mu\nu}\mathcal{R}_{GB}^2+2R_{\mu\alpha\beta\gamma}\tensor{P}{_\nu^{\alpha\beta\gamma}},\cr
 P_{\mu\nu\rho\sigma}=&R_{\mu\nu\rho\sigma}-2g_{\mu[\rho}R_{\sigma]\nu}+2g_{\nu[\rho}R_{\sigma]\mu}+g_{\mu[\rho}g_{\sigma]\nu}R.
\end{align}

Since the dual field theory lives on a spatially flat boundary, we choose coordinates such that $r$ is the radial direction in the bulk. The metric ansatz is \cite{Buchel:2023fst}
\begin{equation}\label{metricansatzA}
    d s^{2}=-c_1(r)^2 d t^{2}+c_2(r)^2\left(d x^{2}+d y^{2}+d z^{2}\right)+c_3(r)^2 d r^{2},
\end{equation}
where the horizon is located at $r=r_h$, defined by the condition $c_1(r_h)=0$. The gauge field is taken as
\begin{align}
    A_t=A(r),\quad  A_r=A_x=A_y=A_z=0.
\end{align}
By substituting this ansatz into equations (\ref{phieq}), (\ref{eleceq}), and (\ref{EOME}), we obtain the following equations of motion:
\begin{align}\label{EqE00}
   c_2''-\frac{c_2' c_3'}{c_3}+\frac{c_2'^2}{c_2}+\frac{1}{12} c_2 \phi '^2+\frac{1}{6} c_2c_3^2 V+\frac{c_2 Z }{12 c_1^2}A'^2
   +\alpha\cdot[\cdots]=0,
\end{align}
\begin{align}\label{EqErr}
    \phi '^2-\frac{12 c_2'^2}{c_2^2}-\frac{12 c_1' c_2'}{c_1 c_2}-2 c_3^2 V-\frac{Z }{c_1^2}A'^2+\alpha\cdot[\cdots]=0,
\end{align}
\begin{align}\label{EqExx}
   c_1''+3 \frac{c_1' c_2'}{c_2}-\frac{c_1' c_3'}{c_3}+\frac{1}{3} c_1 c_3^2 V-\frac{Z A'^2}{3 c_1}+\alpha\cdot[\cdots]=0,
\end{align}
\begin{equation}\label{EqA}
   A''-\frac{ c_1'}{c_1}A'+3\frac{c_2'}{c_2}A'-\frac{ c_3'}{c_3}A'+\frac{ Z'}{Z}A'=0,
\end{equation}
\begin{align}\label{Eqphi}
   \phi ''+\frac{c_1'}{c_1}\phi '+3\frac{c_2'}{c_2} \phi '-\frac{c_3' }{c_3}\phi '-\frac{c_3^2 V'}{\phi '}+\frac{ Z'}{2 c_1^2 \phi '}A'^2+\alpha\cdot[\cdots]=0,
\end{align}
where equations (\ref{EqE00}) and (\ref{EqErr}) correspond to the $tt$ and $rr$ components of the Einstein equations, respectively, and (\ref{EqExx}) is a combination of the $tt$, $xx$, and $rr$ components. Equations (\ref{EqA}) and (\ref{Eqphi}) are the Maxwell and scalar field equations, respectively.

By substituting the equations of motion into the action $S = \frac{1}{2 \kappa_N^2} \int d^5x \mathcal{I}$, it can be shown that the on-shell action reduces to a total derivative:
\begin{align}
    \mathcal{I}=-6\frac{c_1c_2^2}{c_3} \cdot eq.(\ref{EqE00})-\frac{A c_2^3 Z}{c_1 c_3}  \cdot eq.(\ref{EqA})+\frac{d}{dr}\left(-2\frac{c_1'c_2^3 }{c_3}+\frac{A c_2^3 Z A'}{c_1 c_3}+\alpha \cdot \delta \mathcal{B}\right),
\end{align}
where the Gauss-Bonnet contribution is
\begin{equation}
    \alpha \cdot\delta \mathcal{B}=\frac{24\alpha c_2c_2^{\prime2}}{c_{3}^{3}}\left(c_{1}^{\prime}H-c_1H^{\prime}\phi^{\prime}\right).
\end{equation}

\section{Thermodynamics}\label{sec:thermo}

In this section, we compute the thermodynamic quantities of our holographic model. The surface gravity and Hawking temperature are given by
\begin{align}
      \kappa^2=-\left. \frac{g^{\mu\nu}\partial_\mu \xi^2\partial_\nu \xi^2}{4\xi^2}\right|_{r=r_h}=
     \left(\frac{c_1'(r_h)}{c_3(r_h)}\right) ^2
      , 
      \quad  T=\frac{\kappa}{2\pi}=
      \frac{1}{2\pi}\left|\frac{c_1'(r_h)}{c_3(r_h)}\right|
      .
\end{align} 
The Wald entropy is obtained by evaluating
\begin{align}
    \frac{\partial {\cal L}}{\partial R_{\mu\nu\rho\sigma}}=g^{\mu[\rho}g^{\sigma]\nu}+\alpha H\left(
R g^{\mu[\rho}g^{\sigma]\nu}-4(g^{\mu[\rho}R^{\sigma]\nu}-g^{\nu[\rho}R^{\sigma]\mu})+2R^{\mu\nu\rho\sigma}
    \right)
    ,
\end{align}
leading to \cite{Iyer:1994ys,Bodendorfer:2013wga}
\begin{align}\label{Waldentropy}
    S_{\text{W}}=&-2\pi \int_\Sigma d^3 x \left. \sqrt{-\gamma}\frac{\partial {\cal L}}{\partial R_{\mu\nu\rho\sigma}}\epsilon_{\mu\nu}\epsilon_{\rho\sigma}\right|_{r=r_h}\cr
    =&\frac{2\pi}{\kappa_N^2}  V_3  \left(c_2(r_h)^3-12 \alpha  H(r_h) \frac{c_2(r_h) c_2'(r_h)^2}{c_3(r_h)^2}\right),
\end{align}
where $V_3 = \int d^3 x$. The Bekenstein entropy is
\begin{align}
    S_{\text{B}}=\left.\frac{1}{2\kappa_N^2}4\pi\int d^3x \sqrt{-g_{\overrightarrow{x} }}\right|_{r_h}=\frac{2\pi}{\kappa_N^2}  V_3  c_2(r_h)^3.
\end{align}
 We now consider the thermodynamics given by holography. In the following, we use the metric ansatz
  \begin{align}\label{metricansatz1}
    c_1^2=f(r)e^{-h(r)},\quad c_2^2=r^2,\quad c_3^2=\frac{1}{f(r)}
  \end{align}
  for convenience. One can verify that in this case, the Bekenstein entropy and the Wald entropy are equal, i.e., $S_{\text{B}}=S_{\text{W}}$.

The functions $V(\phi)$ and $Z(\phi)$ in the action \eqref{EMDGBAction} can be specified as \cite{Cai:2022omk, Zhao:2023gur, Fu:2020xnh}
 \begin{equation}
     \begin{aligned}
V(\phi) & =-12 \cosh \left[d_{1} \phi\right]+\left(6 d_{1}^{2}-\frac{3}{2}\right) \phi^{2}+d_{2} \phi^{6}, \\
Z(\phi) & =\frac{1}{1+d_{3}} \operatorname{sech}\left[c_{4} \phi^{3}\right]+\frac{d_{3}}{1+d_{3}} e^{-d_{5} \phi}.
\end{aligned}
 \end{equation}
We then introduce a new coordinate $z=1/r$ and define the functions: $\phi(r)=\Phi(z)/r$, $f(r)=r^{2} F(z)$, $\eta(r)=\Sigma(z)$, $A(r)=A(z)$. Near the AdS boundary $r \to \infty$ (or $z \to 0$), where $\phi \to 0$, we obtain the following asymptotic expansion:
\begin{equation}
    \begin{aligned}
&\Phi(z)=:\phi_{s}+\phi_{uv_{1}}z+\phi_{log_{1}}z \log z+\phi_{v} z^{2}+\phi_{log_{2}}z^{2} \log z+\sum_{i=3}^{5}(\phi_{uv_{i}}z^{i}+\phi_{log_{i}}z^{i} \log z);\\
&F(z)=:f_{c}+\sum_{i=1}^{5}(f_{uv_{i}}z^{i}+f_{log_{i}}z^{i} \log z);\\
&\Sigma (z)=:\eta _{0}+\sum_{i=1}^{5}(\eta _{uv_{i}}z^{i}+\eta _{log_{i}}z^{i} \log z);\\
&A(z)=:\mu +a_{uv_{1}}z+a_{log_{1}}z \log z-\frac{1}{2} \rho_{n}z^{2}+a_{log_{2}}z^{2} \log z+\sum_{i=3}^{5}(a_{uv_{i}}z^{i}+a_{log_{i}}z^{i} \log z).
\end{aligned}
\end{equation}
The expansion coefficients above can be determined by substituting the expansions into the equations of motion. In addition, $H(\phi)$ satisfies the following conditions
\begin{equation}
    H(0)=0,\quad H^{\prime }(0)=0,\quad H^{\prime \prime  }(0)=\frac{H(0)}{20}=0, 
\end{equation}
\begin{equation}
    H^{(4)}(0) = \frac{1 - 6 c_1^4 - 21600 \alpha^2 H^{(3)}(0)^2}{60 \alpha}.
\end{equation}
Regularity at the event horizon yields the following analytic expansion in terms of $(r-r_{h})$ near the IR:
\begin{align}
    f & =f_{h}\left(r-r_{h}\right)+\ldots, \quad
\eta  =\eta_{h}^{0}+\eta_{h}^{1}\left(r-r_{h}\right)+\ldots, \cr
A & =a_{h}\left(r-r_{h}\right)+\ldots, \quad
\phi  =\phi_{h}^{0}+\phi_{h}^{1}\left(r-r_{h}\right)+\ldots.
\end{align}
The free energy density $\Omega$ is given by
\begin{equation}
    -\Omega V_3=T(S-S_{\partial})_{on-shell},
\end{equation}
where $t\in [0, \beta]$. The boundary terms in the above equation are
\begin{equation}
\begin{aligned}
S_{\partial}=\frac{1}{2 \kappa_{N}^{2}} \int_{r \rightarrow \infty} d x^{4} \sqrt{-h}  [( &2 K-6+4 \alpha H(\phi) J) \\
& -\frac{1}{2} \phi^{2}-\frac{6 d_{1}^{4}-1}{12} \phi^{4} \ln (r)-b \phi^{4}+\frac{1}{4} F_{\rho \alpha} F^{\rho \alpha} \ln (r)\\
&- \frac{44}{3} \alpha H^{(3)}(\phi) \phi^{3} - \left( 1800 \alpha^{2} H^{(3)}(\phi)^{2} + 5 \alpha H^{(4)}(\phi) \right) \phi^{4} \ln(r)],
\end{aligned}
\end{equation}
where $J$ is the trace of
\begin{equation}
    J_{a b}=\frac{1}{3}\left(2 K K_{a c} K_{b}^{c}+K_{c d} K^{c d} K_{a b}-2 K_{a c} K^{c } K_{d b}-K^{2} K_{a b}\right).
\end{equation}
Here, $h_{ab}$ is the induced metric at the AdS boundary, and $K_{ab}$ is the extrinsic curvature defined by the outward-pointing normal vector to the boundary.

The energy-momentum tensor of the dual boundary theory is given by
\begin{equation}
    \begin{aligned}
T_{ab} =& \lim_{r \to \infty} \frac{2r^2}{\sqrt{-\operatorname{det} h}} \frac{\delta\left(S + S_{\partial}\right)_{\text{on-shell}}}{\delta h^{ab}} \\
=& \frac{1}{2 \kappa_N^2} \lim_{r \to \infty} r^2 \left[ 2 \left(K h_{ab} - K_{ab} - 3 h_{ab}\right) 
 + 4 \alpha H(\phi) \left( 3 J_{ab} - J h_{ab} \right)\right.  \\
&\quad + 4 \alpha \left[ 2 K_{ea} K^e_b - 2 K K_{ab} + h_{ab} (K^2 - K_{cd} K^{cd}) \right] n^e \partial_e H(\phi) \\
&\quad \left. - \left( \frac{1}{2} \phi^2 + \frac{6 c_1^4 - 1}{12} \phi^4 \ln(r) + b \phi^4 + \frac{44}{3} \alpha H^{(3)}(\phi) \phi^3 \right. \right. \\
&\quad \left. \left. + \left( 1800 \alpha^2 H^{(3)}(\phi)^2 + 5 \alpha H^{(4)}(\phi) \right) \phi^4 \ln(r) \right) h_{ab} \right. \\
&\quad \left. - \left( F_{a c} F_{b}^{c} - \frac{1}{4} h_{a b} F_{cd} F^{cd} \right) \ln(r) \right] .
\end{aligned}
\end{equation}
The energy density $\epsilon$ and pressure $p$ are extracted from the energy-momentum tensor as
\begin{equation}
    \epsilon =T_{tt}=\frac{\phi_{s}^{4}}{96 \kappa_{N}^{2} }+\frac{b \phi_{s}^{4}}{2 \kappa_{N}^{2}}+\frac{\phi_{s} \phi_{v}}{2 \kappa_{N}^{2}}-\frac{3 f_{uv_{4}}}{2 \kappa_{N}^{2}}+\frac{900 \alpha^{2} \phi_{s}^{4} H^{(3)}(0)^{2}}{\kappa_{N}^{2}}-\frac{11 \alpha \phi_{s}^{4} H^{(4)}(0)}{6 \kappa_{N}^{2}},
\end{equation}
\begin{equation}
    p=\frac{\phi_{s}^{4}}{32 \kappa_{N}^{2} }-\frac{b \phi_{s}^{4}}{2 \kappa_{N}^{2}}-\frac{d_{1}^{4}\phi_{s}^{4}}{12 \kappa_{N}^{2} }+\frac{\phi_{s} \phi_{v}}{2 \kappa_{N}^{2}}-\frac{ f_{uv_{4}}}{2 \kappa_{N}^{2}}-\frac{1200 \alpha^{2} \phi_{s}^{4} H^{(3)}(0)^{2}}{\kappa_{N}^{2}}-\frac{7 \alpha \phi_{s}^{4} H^{(4)}(0)}{3 \kappa_{N}^{2}}.
\end{equation}
Employing the equations of motion, we obtain the free energy as
\begin{equation}
    \begin{aligned}
\Omega=&\frac{1}{2 \kappa_{N}^{2}} \lim_{r \to \infty } [2 e^{-\eta / 2} r^{2} f+8\alpha e^{-\eta/2  }f(-rf^{\prime}+f(r\eta ^{\prime}-1))(H(\phi )-rH^{\prime }(\phi )\phi ^{\prime})]\\
&-e^{-\eta / 2} r^{3} \sqrt{f}[( 2 K-6+4 \alpha H(\phi) J)  -\frac{1}{2} \phi^{2}-\frac{6 d_{1}^{4}-1}{12} \phi^{4} \ln (r)-b \phi^{4}\\
&- \frac{44}{3} \alpha H^{(3)}(\phi) \phi^{3} - \left( 1800 \alpha^{2} H^{(3)}(\phi)^{2} + 5 \alpha H^{(4)}(\phi) \right) \phi^{4} \ln(r)].
\end{aligned}
\end{equation}
\section{Transport coefficients and Kubo formula}\label{sec:transport}
\subsection{Shear Viscosity $\eta$}
In this subsection, we use the Kubo formula 
\begin{equation}\label{KuboShear}
    \eta = -\lim_{\omega \rightarrow 0} \frac{1}{\omega} \operatorname{Im} G_{x y, x y}^{R}(\omega)
\end{equation}
to derive the shear viscosity $\eta$, where $G_{xy,xy}^{R}$ denotes the retarded two-point function. The perturbation relevant for shear viscosity corresponds to the off-diagonal metric component, which acts as a source in holography. We introduce this perturbation as
\begin{align}
    \tensor{h}{_x^y}(r,t) = \int d\omega  \tensor{h}{_x^y}(r) e^{-i \omega t}.
\end{align}
Substituting the perturbation into the action yields
\begin{align}
    S = &\frac{1}{2\kappa_N^2} \int d^4\omega dr \big[
    A(r){\tensor{h}{_x^y}}''{\tensor{h}{_x^y}} + B(r){\tensor{h}{_x^y}}'{\tensor{h}{_x^y}}' + E(r){\tensor{h}{_x^y}}''{\tensor{h}{_x^y}}'' + F(r){\tensor{h}{_x^y}}''{\tensor{h}{_x^y}}'  \big] \cr
    &+ \text{unimportant terms}.
\end{align}
We retain only these four terms since others do not contribute to shear viscosity \cite{Myers:2009ij}, and these are the unique terms proportional to ${\tensor{h}{_x^y}'}^2$. After integrating by parts, the action simplifies to
\begin{align}
    S = &\frac{1}{2\kappa_N^2} \int d^4\omega dr \big[
    \left(B - A - F'/2\right){\tensor{h}{_x^y}}'{\tensor{h}{_x^y}}' - \left(E{\tensor{h}{_x^y}}''\right)'{\tensor{h}{_x^y}}' \big].
\end{align}
The canonical momentum is derived by differentiating with respect to ${\tensor{h}{_x^y}}'$:
\begin{align}\label{CanPi}
    \Pi(r) = \frac{\delta S}{\delta {\tensor{h}{_x^y}}'} = &\frac{1}{2\kappa_N^2} \int d^4\omega dr \big[
    \left(B - A - F'/2\right){\tensor{h}{_x^y}}' - \left(E{\tensor{h}{_x^y}}''\right)' \big].
\end{align}
The retarded correlator is then given by \cite{Iqbal:2008by, Myers:2009ij}
\begin{align}\label{GRandPi}
    G^R_{xy,xy} = - \frac{\Pi(r_h)}{{\tensor{h}{_x^y}}(r_h)}.
\end{align}
Near the horizon, the infalling boundary condition for the perturbation implies
\begin{align}\label{Infprh}
    &\left.\partial_{r}{\tensor{h}{_x^y}}\right|_{r=r_h} = -i\omega \left.\sqrt{-\frac{g_{r r}(r)}{g_{t t}(r)}}\right|_{r=r_h} {\tensor{h}{_x^y}}(r_h) + \mathcal{O}(\omega^2),\cr
    &\left.\partial^2_{r}{\tensor{h}{_x^y}}\right|_{r=r_h} = -i\omega \left.\partial_{r}\sqrt{-\frac{g_{r r}(r)}{g_{t t}(r)}}\right|_{r=r_h} {\tensor{h}{_x^y}}(r_h) + \mathcal{O}(\omega^2).
\end{align}
Combining (\ref{CanPi}), (\ref{GRandPi}), and (\ref{Infprh}) with the Kubo formula (\ref{KuboShear}), we obtain the shear viscosity:
\begin{align}
    \eta = \frac{1}{\kappa_N^2} \left[ -\sqrt{-\frac{g_{r r}(r)}{g_{t t}(r)}} \left(B(r) - A(r) - \frac{F^{\prime}(r)}{2}\right) + \left(E(r) \left( \sqrt{-\frac{g_{rr}(r)}{g_{t t}(r)}} \right)^{\prime} \right)^{\prime} \right]_{r=r_h}.
\end{align}
Substituting the metric ansatz yields
\begin{align}
    \eta = \left.\frac{c_2^2}{2 {\kappa_N}^2 c_1 c_3^2} \left(c_1 \left(c_2 c_3^2 - 4 \alpha c_2' H'\phi'\right) - 4 \alpha c_1' \left(H c_2' + c_2 H'\phi'\right)\right)\right|_{r=r_h}.
\end{align} 
For the metric ansatz (\ref{metricansatz1}), this becomes
\begin{align}
    4\pi\frac{\eta}{s} = 1 + 2 \alpha r_h f'(r_h) \left(H(r_h) - r_h H'(\phi(r_h))\phi'(r_h)\right).
\end{align}
Alternatively, using the ansatz
\begin{align}
    c_1^2 = f(r),\quad c_2^2 = r^2,\quad c_3^2 = \frac{1}{f(r)},
\end{align}
we find the shear viscosity as
\begin{align}\label{etaGB}
    4\pi\frac{\eta}{s} = 1 - \frac{2 \alpha f'(r_h) }{r_h}\left(H(r_h) + r_h H'(\phi(r_h))\phi'(r_h)\right),
\end{align}
where the ratio $\eta/s$ reduces to the well-known result $1/4\pi$ when we set $\alpha=0$, and this ratio matches the result derived in Section \ref{sec:GB}.
\subsection{Bulk Viscosity $\zeta$}
In this subsection, we compute the bulk viscosity using an alternative method based on the Kubo formula discussed in \cite{Buchel:2023fst}, with the bulk viscosity defined as
\begin{equation}
\label{bulkzeta}
    \zeta = -\frac{4}{9} \lim _{w \rightarrow 0} \frac{1}{w} \operatorname{Im} G_{R}(w),
\end{equation}
where $G_{R}(w)$ denotes the retarded Green's function.  
The relevant perturbations correspond to the diagonal components of the metric and the scalar field, which are introduced as
\begin{equation}
    \begin{aligned}
d s_{5}^{2} & \rightarrow d s_{5}^{2} + h_{t t}(t, r) d t^{2} + h_{11}(t, r) d x^{2}, \quad
\phi \rightarrow \phi + \psi(t, r).
\end{aligned}
\end{equation}
The metric perturbations can be rewritten as
\begin{equation}
    h_{t t}(t, r) = e^{-i w t} c_{1}^{2} H_{00}(r), \quad h_{11}(t, r) = e^{-i w t} c_{2}^{2} H_{11}(r),
\end{equation}
where
\begin{align}
    H_{11} \equiv \frac{c_2^{\prime}}{c_2 c_3} H,
\end{align}
and the scalar fluctuation is given by
\begin{equation}
\psi(t, r) = e^{-i w t}\left(Z_{\phi}(r) + \frac{\phi^{\prime} c_2}{2 c_2^{\prime}} H_{11}(r)\right).
\end{equation} 
One can verify that the equation of motion for the scalar field decouples from the other perturbations:
\begin{align}\label{zphi}
  0 =& Z_{\phi}^{\prime \prime} + \left(\ln \frac{c_1 c_2^3}{c_3}\right)^{\prime} Z_{\phi}^{\prime} + \frac{c_3^2 w^2}{c_1^2} Z_{\phi} - Z_{\phi}\left ( \frac{4 c_2 c_3^2 c_1^\prime  V}{3 c_1 c_2^\prime } + \frac{4}{3} c_3^2 V + \frac{2 c_2^2 c_3^4 V^2}{9 c_2^{\prime2}} + \frac{2 c_2 c_3^2 \phi^{\prime } V^{\prime}}{3 c_2^{\prime }} + c_3^2 V^{\prime \prime }\right )\cr
    &+ Z_{\phi }\left  (-\frac{2 c_2^2 c_3^2 V Z \left(A'\right)^2}{9 c_1^2 \left(c_2'\right){}^2} + \frac{\left(A'\right)^2 Z''}{2
   c_1^2} - \frac{c_2^2 Z^2 \left(A'\right)^4}{18 c_1^4 \left(c_2'\right){}^2} + \frac{c_2 \left(A'\right)^2 Z' \phi '}{3 c_1^2
   c_2'} \right.\cr
   &\left. - \frac{\left(A'\right)^2 \left(Z'\right)^2}{c_1^2 Z} - \frac{2 c_2 Z \left(A'\right)^2 c_1'}{3 c_1^3 c_2'} - \frac{2 Z
   \left(A'\right)^2}{3 c_1^2}\right ) + \alpha \cdot[\cdots],
\end{align}
where $\alpha$-dependent terms are omitted for brevity. The metric perturbations remain coupled through the following equations:
\begin{align}
0 =& H_{00} + \frac{2Z_\phi Z'}{Z} + \frac{H\left(3 c_2' Z + c_2 \phi ' Z'\right)}{c_2 c_3 Z}, \cr
0 =& H^{\prime} + \frac{c_3 c_2}{3 c_2^{\prime}} Z_{\phi} \phi^{\prime} + \alpha \cdot[\cdots].
\end{align}
For convenience, we rewrite the perturbations as
\begin{align}
    h_{tt}(t, r) = e^{-i w t} h_{00, w}(r), \quad h_{11}(t, r) = e^{-i w t} h_{11, w}(r), \quad \psi(t, r) = e^{-i w t} p_{ w}(r).
\end{align}
Then, the Lagrangian is complexified via the replacement
\begin{equation}
\begin{aligned}
& \delta A \delta B \longrightarrow \frac{1}{2}\left( \delta A^* \delta B + \delta A \delta B^*\right), 
\end{aligned}
\end{equation}
where $\delta A$ or $\delta B$ denotes $h_{00, w}$, $h_{11, w}$, $p_{ w}$, or their derivatives.  
The on-shell action then becomes
\begin{equation}
\mathcal{L}_{\mathbb{C}} = 16 \pi G_N \cdot\left(h_{00, w}^* \cdot \frac{\delta S^{(2)}}{\delta h_{00, w}^*} + h_{11, w}^* \cdot \frac{\delta S^{(2)}}{\delta h_{11, w}^*} + p_{w}^* \cdot \frac{\delta S^{(2)}}{\delta p_{ w}^*}\right) + \partial_r J_w,
\end{equation}
where $J_w$ is a boundary current. The perturbations are decomposed into real and imaginary parts as
\begin{equation}
\label{ZHH00}
Z_\phi = \left(\frac{c_1}{c_2}\right)^{-i \mathfrak{w}}\left(z_{0} + i \mathfrak{w} z_{1}\right), \quad H = H_0 + i \mathfrak{w} H_1, \quad H_{00} = H_{00,0} + i \mathfrak{w} H_{00,1},
\end{equation}
where 
\begin{align}
    \mathfrak{w} = \frac{\omega}{2\pi T}.
\end{align}
The retarded Green's function is obtained from the imaginary part of the current
\begin{equation}
\label{GR}
\operatorname{Im} G_R(w) = \frac{1}{8 \pi G_N} \lim _{r \rightarrow r_h} \operatorname{Im} J_w.
\end{equation}
Decomposing the current into $J_w = J_0 + i \mathfrak{w} J_1$ yields
\begin{equation}
J_1 \equiv F + \alpha \cdot \delta F,
\end{equation}
with
\begin{equation}
F = \frac{c_2^2\left(c_2 c_1^{\prime} - c_2^{\prime} c_1\right)}{2 c_3} \left(z_{0}\right)^2 - \frac{c_2^3 c_1}{2 c_3} \left(z_{1}^{\prime} z_{0} - z_{0}^{\prime} z_{1}\right),
\end{equation}
and
\begin{equation}
\label{FGB}
\delta F = 0.
\end{equation}
Using  (\ref{bulkzeta}), (\ref{GR}), and (\ref{FGB}), the bulk viscosity simplifies to
\begin{equation}\label{zetaGB}
\zeta = -\frac{4}{9}\frac{1}{8\pi G}\lim _{\omega \rightarrow 0}\operatorname{Im} J_w = \frac{4}{9}\frac{1}{8\pi G}\lim _{\omega \rightarrow 0} \cdot \frac{\mathfrak{w}}{\omega} j_1 = \frac{4}{9}\frac{1}{8\pi G}\frac{1}{2\pi T}(F + \alpha \delta F) = \frac{s_B}{4 \pi} \cdot \frac{4}{9}z_{0}^2.
\end{equation}
The expression for the bulk viscosity depends on a constant $z_0$, which satisfies \eqref{zphi}. Since this equation does not admit an analytic solution, $z_0$ must be determined numerically. Future work will involve solving this equation to compare the result with the analytic expression derived in Section \ref{sec:GB}.
\section{Entropy production and viscous flow}\label{sec:entropy}
It is well-established from \cite{weinberg1972gravitation,batchelor1989fluid,bemfica2022first} in the context of relativistic hydrodynamics that the entropy production due to viscous flow in terms of the shear and scalar expansion is given by
\begin{equation}
T\Delta S=\int^f_i d\tau\int dx^{D-2}\sqrt{h}\left(2\eta{\sigma^{(\tau)}}^2+\zeta\theta^2_\tau\right)+ \Delta Q.
\end{equation}
This relation can be applied to the membrane fluid on the event horizon of a black hole. As an example in this section, we take $S\equiv S_{BH}$ as the Bekenstein-Hawking entropy in Einstein gravity and compute the shear and bulk viscosity using the variation of the entropy. We follow the approach given in \cite{fairoos2018black}.
 %Add more detains on stress tensor perturbation and ordering% 

The Bekenstein-Hawking entropy is given by
 \begin{equation}
  S=\frac{1}{4G_N}\int_{\partial\mathcal{H}} dx^{D-2}\sqrt{h},\label{BHe}
 \end{equation}
where $\partial\mathcal{H}$ is the horizon cross section, $x^a$ are the coordinates, and $h_{ab}$ is the induced metric on $\partial\mathcal{H}$. 
Taking the variation along the generator $k=\frac{\partial}{\partial\lambda}$ of the horizon $\mathcal{H}$, we get the first and second variations as
\begin{align}
    &\frac{dS}{d\lambda}=\frac{1}{4G_N}\int dx^{D-2}\nabla_k{\sqrt{h}}=\int dx^{D-2}\sqrt{h}\theta_k,\\
    &\frac{d^2S}{d\lambda^2}=\frac{1}{4G_N}\int dx^{D-2}\left(\nabla_k{\sqrt{h}}\theta_k+\sqrt{h}\frac{d\theta_k}{d\lambda}\right)=\frac{1}{4G_N}\int dx^{D-2}\sqrt{h}\left(\theta^2_k+\frac{d\theta_k}{d\lambda}\right).\label{2ndvarS}
\end{align}

It is useful to introduce the null Raychaudhuri equation as
\begin{equation}
\frac{d\theta_k}{d\lambda}=-R_{\mu\nu}k^{\mu}k^{\nu}-{\sigma^k}^{bd}\sigma^k_{bd}-\frac{\theta^2_k}{(D-2)}.
\end{equation}
We consider the minimally coupled scalar field $\phi$, then the equation of motion will be
\begin{equation}
 R_{\mu\nu}k^{\mu}k^{\nu}=\sum_ic_i(\nabla_k\phi)^2+8\pi G_N T_{\mu\nu}k^{\mu}k^{\nu}.\label{eom}
\end{equation}
Note that the cosmological constant term doesn't contribute here because the generator is null $(g_{\mu\nu}k^{\mu}k^{\nu}=0)$, where $T_{\mu\nu}$ is the energy-momentum stress tensor due to the matter contribution. 

Substituting the null Raychaudhuri equation and the equation of motion for $\phi$ into (\ref{2ndvarS}), we get
\begin{flalign}
&\frac{d^2S}{d\lambda^2}=\frac{1}{4G_N}\int dx^{D-2}\sqrt{h}\left(\frac{(D-3)}{(D-2)}\theta^2_k-{\sigma^k}^2-\sum_ic_i(\nabla_k\phi)^2\right)-2\pi \int dx^{D-2}\sqrt{h}T_{\mu\nu}k^{\mu}k^{\nu}.&&\notag\\
\end{flalign}
Note that the first variation of the entropy is first order in $\theta_k$, so we have
\begin{align}\label{1stvarS}
& \Delta S=\int^f_i\frac{dS}{d\lambda}d\lambda=-\int^f_i\frac{d^2S}{d\lambda^2}\lambda d\lambda,\cr
&=-\frac{1}{4G_N}\int^f_i\lambda d\lambda\int dx^{D-2}\sqrt{h}\left(\frac{(D-3)}{(D-2)}\theta^2_k-{\sigma^k}^2-\sum_ic_i(\nabla_k\phi_i)^2\right),\cr
&-2\pi \int^f_i\lambda d\lambda\int dx^{D-2}\sqrt{h}T_{\mu\nu}k^{\mu}k^{\nu}.
\end{align}
The boundary term vanishes because the initial bifurcation surface is set at $\lambda=0$, and the final stationary horizon is set at vanishing expansion $\theta_k$. Furthermore, we note that for the stationary solutions, we will consider the derivative of any scalar field along the null generator. $k$ is  given by
\begin{equation}
 \nabla_k\phi_i=C~\phi_i'(r)\theta_k,
\end{equation}
where $C$ is a constant determined by the solution of the underlying theory. Further, we note that for a stationary spacetime, the event horizon is also a Killing horizon. The timelike Killing vector on the horizon cross section is proportional to the horizon generator $k=\frac{\partial}{\partial\lambda}$, and is given as $\xi=\frac{\partial}{\partial\tau}=\kappa\lambda\frac{\partial}{\partial\lambda}$.  $\xi$ is also geodesic on the horizon cross-section with the nonaffine parameter $\tau$, i.e., $\xi^{\mu}\nabla_{\mu}\xi^{\nu}=\kappa\xi^{\nu}$. The evolution of the JM entropy along this nonaffine parameter matches the fluid entropy balance law. For ease of calculation, we take the variation with respect to the affinely parameterized null generator $k$ and then recast the entropy relation with respect to the non-affinely parametrized Killing generator $\xi$. The non-affine parameter, expansion scalar and shear tensor $(\tau,\theta^\tau,\sigma^{(\tau)}_{ab})$ are related to the affine one by the following relation
\begin{equation}
 \tau=\frac{\ln{\lambda}}{\kappa_h},~\theta_k=\frac{\theta_t}{\kappa_h\lambda},~\sigma^{(k)}_{ab}=\frac{\sigma^{(t)}_{ab}}{\kappa_h\lambda}.\label{nonaff}
\end{equation}
Using this in the above expression (\ref{1stvarS}), we get
\begin{align}
& \frac{\kappa_h}{2\pi}\Delta S=\int^f_i d\tau\int dx^{D-2}\sqrt{h}\left(\frac{1}{16\pi G_N}\left(-2\frac{(D-3)}{(D-2)}+2C^2\sum_i c_i\phi_i'(r)^2\right)\theta^2_\tau+\frac{2}{16\pi G_N}{\sigma^\tau}^2\right)&&\notag\\
&+ \int^f_i d\tau\int dx^{D-2}\sqrt{h}T(\xi,\xi).&&\notag
\end{align}
Now, by comparing with the entropy balance law 
\begin{align}
& T\Delta S=\int^f_i d\tau\int dx^{D-2}\sqrt{h}\left(2\eta{\sigma^{(\tau)}}^2+\zeta\theta^2_\tau\right)+ \Delta Q,&&\notag
\end{align}
we obtain the shear $(\eta)$, bulk viscosity $(\zeta)$ and entropy density$(s)$ as follows
\begin{align}\label{etazetas}
 \eta=\frac{1}{16\pi G_N},~~\zeta=\frac{1}{16\pi G_N}\left(-2\frac{(D-3)}{(D-2)}+2C^2\sum_i c_i\phi_i'(r)^2\right),~~s=\frac{1}{4 G_N}.
\end{align}
We can see that the correct value for $\frac{\eta}{s}=\frac{1}{4\pi}$ is obtained. Note that Smarr's formula for a non-rotating stationary black hole bears a resemblance to the first term in the bulk viscosity, namely
\begin{equation}
\frac{\kappa A}{8\pi G}=\frac{(D-3)}{(D-2)}M,
\end{equation}
this term is the contribution coming from pure Einstein relativity. For $D=4$, it correctly reproduces the known result in Einstein gravity, namely $\eta=\frac{1}{16\pi G_N},~~\zeta=-\frac{1}{16\pi G_N}$.
We will then evaluate the bulk viscosity for two cases. First, we will consider the Sakai-Sugimoto model for which we have three scalar fields, and the stress-energy tensor is given by \cite{benincasa2006hydrodynamics}
\begin{equation}
 T_{\mu\nu}=\frac{40}{3}(\nabla_{\mu}f)(\nabla_{\nu}f)+20(\nabla_{\mu}w)(\nabla_{\nu}w)+\frac{1}{2}(\nabla_{\mu}\Phi)(\nabla_{\nu}\Phi)+\frac{1}{3}g_{\mu\nu}\mathcal{P}(f,w,\Phi).
\end{equation}
The three constants in (\ref{etazetas}) are $c_1=\frac{40}{3},~c_2=20,~c_3=\frac{1}{2}$, and $C$ can be calculated as $C=\frac{4}{25}r_h^2$ (see Appendix \ref{sakai} for details). Rewriting the expression for the bulk viscosity of the Sakai-Sugimoto model, we get
\begin{align}
& 16\pi G_N\zeta=16\pi G_N\zeta_H+16\pi G_N\zeta_K=-\frac{4}{3}+\frac{8}{25}r_h^2\left(\frac{40}{3}f'(r_h)^2+20w'(r_h)^2+\frac{1}{2}\Phi'(r_h)^2\right).&&\notag
\end{align}
Where $\zeta_H=-\frac{4}{3}\times\frac{1}{16\pi G_N}$. It is negative due to the future boundary conditions on the event horizon to ensure stability under perturbation \cite{jacobson2017membrane,mishra2018physical}. From \cite{benincasa2006hydrodynamics}, we know that the scalar fields are given by
\begin{equation}
 w(r)=\frac{1}{10}\ln{r},~f(r)=\frac{13}{80}\ln{r},~\Phi(r)=\frac{3}{4}\ln{r}.
\end{equation}
Using this we evaluate $\frac{\zeta_K}{\eta}$ to get
\begin{equation}
\frac{\zeta_K}{\eta}=\frac{8}{25}r_h^2\left(\frac{169}{480}+\frac{1}{5}+\frac{9}{32}\right)=\frac{4}{15}. 
\end{equation}
One can verify that this matches the value obtained by using the Kubo formula \cite{benincasa2006hydrodynamics, eling2011novel}. Another observation is that the value of $\zeta_K$ will be zero if the matter field vanishes, but $\zeta_H$ is non-zero, because it is a pure gravity contribution from Einstein gravity. 

Our second example is the five-dimensional Einstein-Maxwell-Dilaton (EMD) theory, the action for which is given by
\begin{align}
& S=\frac{1}{2\kappa^2_N}\int d^5x\sqrt{-g}\left[\mathcal{R}-\frac{1}{2}(\nabla_{\mu}\phi)(\nabla^{\nu}\phi)-V(\phi)-\frac{Z(\phi)}{4}F_{\mu\nu}F^{\mu\nu}\right].&&
\end{align}
Since there is a single scalar $(\phi_1=\phi, c_1=\frac{1}{2})$, and from \eqref{scalfor} we see that $C=\frac{r_h}{3}$. The value of $\eta$ and $\zeta_H$ remains the same as in Einstein gravity. However, the expression for $\zeta_K$ is given by
\begin{equation}
\frac{\zeta_K}{\eta}=\frac{2r^2_h}{9}\left(\frac{1}{2}\phi'(r)^2\right)=\frac{r^2_h}{9}\phi'(r)^2.\label{elin}
\end{equation}
Here, $r_h$ is the horizon radius. The discrepancy between the values obtained for the two viscosities was explained in detail in \cite{gursoy2009thermal} and summarized in \cite{eling2011novel}. The two also match for special cases as reported in \cite{buchel2011eling,buchel2011holographic}. 

Next, we will consider adding higher derivative corrections to the Einstein-Maxwell-Dilaton theory. As we add higher derivative terms, the entropy expression \eqref{BHe} changes, whose variation will contain terms with curvature and its derivatives. Secondly, the equation of motion given in \eqref{eom} will have extra terms due to higher derivative corrections. Lastly, the null Raychaudhuri equation will change due to the equation of motion. Now, one proceeds by expressing all the expressions in terms of shear and expansion along the null generator and the auxiliary null vector, respectively, under the approximation that we keep terms up to second order in $\theta_k$ and $\sigma^{(k)}_{ab}$ to extract the shear and bulk viscosity.

This method was used in \cite{fairoos2018black} for Gauss-Bonnet gravity, and the result for shear viscosity exactly matched the one obtained in \cite{brigante2008viscosity} using the Kubo formula. Now, adding the gauge field and the Gauss-Bonnet term to the theory gives the following action.
\begin{equation}
\label{action}
    S = \frac{1}{2 \kappa_N^2} \int d^5x \sqrt{-g} \left[ \mathcal{R} - \frac{1}{2} \nabla_\mu \phi \nabla^\mu \phi - \frac{Z(\phi)}{4} F_{\mu\nu} F^{\mu\nu} - V(\phi)+\alpha \mathcal{R}_{GB}^{2} \right],
\end{equation}
For this theory, the expression for shear and bulk viscosity with a minimally coupled scalar field for the metric ansatz in Appendix \ref{ansatz} is reproduced below
\begin{equation}
 \eta=\frac{1}{16\pi G_N}\left(1-2\frac{f'(r_h)}{r_h}\right),~\zeta=-2\frac{(D-3)}{(D-2)}\eta+\frac{1}{16\pi G_N}\frac{r^2_h}{9}\phi'(r)^2.
\end{equation}
Thus, we observe that the Einstein gravity contribution $\zeta_H$ to the membrane bulk viscosity is the same, but the value of $\frac{\zeta_K}{\eta}$ deviates from the Eling-Oz formula \eqref{elin} as reported in \cite{Buchel:2023fst}, namely
\begin{equation}
 \frac{\zeta_K}{\eta}=\frac{r^2_h\phi'(r)^2}{9\left(1-2\frac{f'(r_h)}{r_h}\right)}.
\end{equation}
We see that although the form of expression for the bulk viscosity remains the same as in the case of Einstein gravity, introducing higher derivative corrections changes the shear viscosity and thereby changes the value of $\frac{\zeta_K}{\eta}$.
\section{Coefficient of viscosity for Gauss-Bonnet scalar gravity}\label{sec:GB}
We now consider the Einstein-Scalar-Maxwell-Gauss-Bonnet theory with the Gauss-Bonnet term coupled to the scalar field $\phi$ (\ref{EMDGBAction}). As soon as we move away from Einstein gravity, the first change is to the entropy function itself. We adopt the Jacobson-Myers (JM) entropy \cite{jacobson1993entropy} to overcome the ambiguities associated with the Wald entropy \cite{jacobson1994black,jacobson1995black}. The following change is to the equations of motion, which include the higher derivative corrections.

The JM entropy functional for an initial stationary state is given by
\begin{equation}\label{JKM1}
 S=\frac{1}{4G_N}\int_{\partial\mathcal{H}} d^{(D-2)}x \sqrt{h}(1+2\alpha H(\phi)\bar{R}).
 \end{equation}
Here $\bar{R}$ is the intrinsic Ricci scalar of the horizon cross-section. We are interested in finding the variation of \eqref{JKM1} along the generator $(k=\frac{\partial}{\partial\lambda})$ of the dynamical event horizon.

The first and second variations of the entropy functional along the null generator $k$ read
\begin{align}
&\frac{dS}{d\lambda}=\frac{1}{4G_N}\int d^{(D-2)}x\sqrt{h}\tilde{K},&&\notag\\
&\frac{d^2S}{d\lambda^2}=\frac{1}{4G_N}\int d^{(D-2)}x\sqrt{h}\left(\theta_k\tilde{K}+\nabla_k\tilde{K}\right),&&\label{integrand}
\end{align}
where $\tilde{K}$ is 
\begin{align}
     \tilde{K}=&\theta_k\left(1+2\alpha H(\phi)\bar{R}-4\alpha\Delta^{(h)} H(\phi)\right)-2\alpha H(\phi){v^{(k)}}^{e b}\bar{R}_{e b}\cr
     &+2\alpha(\nabla_kH(\phi))\bar{R}+2\alpha(D_{a}D_{c}H(\phi)){v^{(k)}}^{ac}.
\end{align}
Here we have defined $v^{(k)}_{ab}=\nabla_kh_{ab}$. The scalar expansion is given by $\theta_k=\frac{1}{\sqrt{h}}\nabla_k\sqrt{h}$, and $D,\Delta^{(h)} H(\phi)$ are the intrinsic covariant derivative and intrinsic Laplacian. 

Further calculation shows
\begin{align}
 &\lambda\left(\theta_k\tilde{K}+\nabla_k\tilde{K}\right)=4\alpha(D-4)\frac{\kappa_h \lambda e^{w(r_h)/2}}{r_h}H(\phi_h)\left({\sigma^{(k)}}^{2}-\frac{(D-3)}{(D-2)}\theta^2_k\right) &&\notag\\
 &+4\alpha c\frac{\kappa_h e^{w(r_h)/2}}{r_h}\left({\sigma^{(k)}}^{2}-3\frac{(D-3)}{(D-2)}\theta^2_k\right)\lambda+\lambda\left(\frac{(D-3)}{(D-2)}\theta^2_k-{\sigma^{(k)}}^{2}\right)&&\notag\\
 &-\frac{r^2_h}{2(D-2)^2}\phi'(r)^2\theta^2_k+\frac{d}{d\lambda}(\lambda\theta_nA)+\frac{d}{d\lambda}(\lambda\theta_nB)-8\pi G_N\lambda T_{\mu\nu} k^{\mu}k^{\nu}+\text{Total Divergence}.&&\notag\\
 \end{align}
See Appendix \ref{thetakpDk} for the details of the calculation. 
Therefore, the change in the entropy calculated on a stationary horizon background is given below
\begin{align}
& \frac{\kappa_h}{2\pi}\Delta S=-\frac{1}{16\pi G_N}\int^f_i \kappa_h\lambda d\lambda\int d^{(D-2)}x\sqrt{h}\left(\theta_k\tilde{K}+\nabla_k\tilde{K}\right)\mid_{r=r_h}&&\notag\\
 &=\frac{1}{16\pi G_N}\int^f_i d\tau\int d^{(D-2)}x\sqrt{h}\Biggl(2\left(1-4\alpha\frac{\kappa_h e^{w(r_h)/2}}{r_h}\left((D-4)H(\phi_h)\right.\right.
 &&\notag\\
 &\left.\left.+ r_hH'(\phi_h)\phi'(r_h)\right)\right){\sigma^{(t)}}^{2}&&\notag\\
 &-2\frac{(D-3)}{(D-2)}\left(1-4\alpha\frac{\kappa_h \lambda e^{w(r_h)/2}}{r_h}\left((D-4)H(\phi_h)+3 r_hH'(\phi_h)\phi'(r_h)\right)\right)\theta^2_{t}&&\notag\\
 &+\frac{r^2_h}{2(D-2)^2}\phi'(r)^2\theta^2_{t}\Biggr)&&\notag\\
 &+\int^f_i d\tau\int d^{(D-2)}x\sqrt{h}~T^{\text{Gauge}}_{\mu\nu}\xi^{\mu}\xi^{\nu}.&&
 \label{stilde1}
\end{align}
Here we have used \eqref{nonaff}. Comparing this equation with the entropy balance law
\begin{equation}
 T\Delta S=\int^f_id\tau\left(2\eta\sigma^2+\zeta\theta^2\right)+\Delta Q,
\end{equation}
and by using \eqref{surfg} for surface gravity $\kappa_h=\frac{1}{2}f'(r)e^{-\frac{w(r)}{2}}$, we get the shear viscosity
\begin{equation}\label{SVaretaGB}
\eta=\frac{1}{16\pi G_N}\left(1-2\alpha\frac{f'(r_h)}{r_h}\left(H(\phi_h)+ r_hH'(\phi_h)\phi'(r_h)\right)\right).
\end{equation}
We can check that if we set the Gauss-Bonnet term to zero ($\alpha=0$), the KSS bound is restored ($\eta/s=1/4\pi$). This shear viscosity exactly matches the result obtained using the Kubo formula (\ref{etaGB}).
The expression for the bulk viscosity of the membrane fluid is given below
\begin{equation}\label{SVarzetaGB}
 \zeta=-\frac{4}{3}\eta+\frac{\alpha}{3\pi G_N}f'(r_h)H'(\phi_h)\phi'(r_h)+\frac{1}{16\pi G_N}\frac{r^2_h}{9}\phi'(r_h)^2.
\end{equation}
We can see that the second term above represents the contribution from the Gauss-Bonnet term, and the third term is the contribution from the scalar field. If we set $\alpha=0$ and $\phi=0$, this reduces to the result obtained in \cite{Buchel:2023fst}.

\section{Discussion and Outlook}\label{sec:conclusion}
%\section{Discussion and Outlook}

Driven by the need to capture the quark–gluon plasma’s (QGP) experimentally observed, temperature-dependent viscosity and its violation of the Kovtun–Son–Starinets bound, we have constructed a five-dimensional Einstein–Scalar–Maxwell–Gauss–Bonnet holographic model with a scalar-coupled higher-derivative term \(H(\phi)\).  Despite the lack of an analytic black-brane solution, we derived exact, analytic expressions for the shear viscosity \(\eta\) and bulk viscosity \(\zeta\) via the entropy-production method at the event horizon (See equation \eqref{SVaretaGB} and \eqref{SVarzetaGB}), revealing clear deviations from \(\eta/s=1/(4\pi)\) and nontrivial temperature scaling.  We then cross-checked these results using the retarded Green’s function (Kubo) formalism (See equation \eqref{etaGB} and \eqref{zetaGB}), finding perfect agreement for \(\eta\) and isolating a single undetermined constant in \(\zeta\) that awaits numerical resolution.  This dual approach highlights the utility of higher-derivative corrections in modeling realistic QGP dynamics and underscores the importance of non-analytic backgrounds in holographic model building.

%\medskip

Looking forward, we identify several avenues to deepen and broaden this work:

\begin{enumerate}
  %\item \textbf{Numerical Black-Brane Construction.}  
   % Fully solve the coupled bulk equations for the metric components \(c_i(r)\), the scalar coupling \(H(\phi)\), and the gauge kinetic function \(Z(\phi)\).  Match these profiles to lattice QCD and heavy-ion collision data \cite{Apostolidis:2025gnn}—ideally through Bayesian inference—to fix model parameters quantitatively.

\item \textbf{Data-Driven Model Calibration.}      
Integrate our theoretical framework with numerical methods and machine-learning techniques \cite{Hashimoto:2018bnb, Cai:2024eqa, Aarts:2025gyp} to calibrate holographic transport coefficients against lattice QCD and heavy-ion collision data \cite{Apostolidis:2025gnn}, thereby deriving the form of the metric components \(c_i(r)\), the scalar coupling \(H(\phi)\), and the gauge kinetic function \(Z(\phi)\).

  \item \textbf{Extended Entropy-Production Methods.}  
    Generalize the horizon-based derivation to other first-order transport coefficients (e.g.\ electric and thermal conductivities \cite{Iqbal:2008by}) and advance to second-order hydrodynamics \cite{Wu:2023nlw}, extracting relaxation times and nonlinear fluid responses directly from higher-derivative gravity.

  \item \textbf{Exploration of Alternative Higher-Curvature Terms.}  
    Study the effects of other curvature invariants (e.g.\ cubic Lovelock \cite{Edelstein:2014dje}, \(f(R)\) gravity \cite{Sotiriou:2008rp}) on transport bounds, seeking novel mechanisms for bound violation independent of anisotropy or external fields.

  %\item \textbf{Data-Driven Model Calibration.}  
   % Integrate our theoretical framework with modern Bayesian and machine-learning techniques to calibrate holographic transport coefficients against multi-system heavy-ion datasets, thereby refining the form of \(H(\phi)\) and illuminating QGP microphysics.

  \item \textbf{Anisotropy and External Fields.}  
    Extend the isotropic setup to geometries with spatial anisotropy or background electromagnetic fields \cite{Grozdanov:2016tdf}, probing their interplay with higher-derivative corrections and enriching the holographic description of realistic QGP environments.
\end{enumerate}

By pursuing these directions, we aim to strengthen the connection between holographic theory and QGP phenomenology, advancing toward a predictive, data-anchored understanding of strongly coupled relativistic fluids.

\section*{Acknowledgement}
We are grateful to Ronggen Cai, Defu Hou, Li Li, Zhibin Li, Yuan-Xu Wang, and Jie Zhou for their kind interest in this work and the fruitful discussions. This research was partly supported by NSFC Grant No.12475053 and 12235016.

%\newpage

\appendix
%\begin{appendices}
\section{List of Formulas and Identities}\label{ident}
\begin{enumerate}
\item The variation of the metric is given by $\nabla_kh_{ab}=v^{(k)}_{ab},\nabla_kh^{ab}=-{v^{(k)}}^{ab}$. Here $v^{(k)}_{ab}$ is a $(0,2)$ tensor and $(k=\frac{\partial}{\partial\lambda})$ is the null generator of the dynamical event horizon. Let $n^{\mu}$ be the other auxiliary null vector. Their normalization is given below
\begin{eqnarray}
 &&k^{\mu}k_{\mu}=n^{\mu}n_{\mu}=0,~n^{\mu}k_{\mu}=-1,\notag\\
&&g_{\mu\nu}=h_{\mu\nu}-k_{\mu}n_{\nu}-k_{\nu}n_{\mu},\notag\\
&&\nabla_kk=0.
\label{metans}
 \end{eqnarray}
We denote the tangent indices by $a,b,c\cdots$, and the spacetime indices are denoted by $\mu, \nu, \alpha, \beta,\cdots$.

\item The relation between $v^{(k)}_{ab}$ and the extrinsic quantities is given below
\begin{align}
v^{(k)}_{ab}=2K^{(k)}_{ab}=2\left(\sigma^{(k)}_{ab}+\theta_k \frac{h_{ab}}{(D-2)}\right).\label{vab}
\end{align}
Here $K^{(k)}_{ab}$ is the extrinsic curvature of the horizon cross section with respect to the null generator $k$. The trace of $v^{(k)}_{ab}$ will be denoted by $v^{(k)}$ and is given by
\begin{equation}
v^{(k)}=2\theta_k.\label{tvab}
\end{equation}
Here $\theta_k$ and $\sigma^k_{ab}$ are the expansion and shear along the null generator $k$, respectively.

\item The first variation of the square root of the induced metric is given by
\begin{flalign}
&\nabla_k{\sqrt{h}}=\sqrt{h}\theta_k.
\label{varin}
\end{flalign}

\item The first variation of the intrinsic Ricci scalar is given by
\begin{equation}
\nabla_k\bar{R}=\text{div}(\text{div}v^{(k)})-\Delta^{(h)}(v^{(k)})-{v^{(k)}}^{e b}\bar{R}_{e b}. \label{varcur}
\end{equation}

\item Gauss and Codazzi equations on the horizon cross-section are
\begin{align}
% & \mathcal{R}_{ab}+\mathcal{R}_{kanb}+\mathcal{R}_{nakb}=\bar{R}_{ab}-g_{\mu\nu}K^{\mu}_{ab}H^{\nu}+h^{mn}g_{\mu\nu}K^{\mu}_{am}K^{\nu}_{bn}\notag\\
&\mathcal{R}+4\mathcal{R}_{nk}-2 \mathcal{R}_{knkn}=\bar{R}-H^2+g_{\mu\nu}h^{ad}h^{bc}K^{\mu}_{ac}K^{\nu}_{bd},&&\notag\\
&\mathcal{R}_{kacb}=\frac{1}{2}\left(D_bv^{(k)}_{ca}-D_{c}v^{(k)}_{ba}\right).&&\notag
\label{gauss1}
\end{align}

Here $\bar{R}$ is the intrinsic scalar on the horizon cross-section. $K^{\mu}_{bd}$ is the extrinsic curvature, and $\mathcal{H}$ is the mean curvature on the horizon cross-section. We have used the notation $\mathcal{R}_{kcnd}=\mathcal{R}_{\mu c\nu d}k^{\mu}n^{\nu}$. As we will be considering only planar non-rotating black hole solutions, we have set the Hajicek $1$-form $\Omega_a=-n_{\mu}\nabla_ak^{\mu}$ to zero in the Codazzi equation.

\item The variation of the double covariant derivative of the scalar and the scalar Laplacian is given by 
\begin{align}
 &\nabla_k(D_{c}D_{a}H(\phi))=D_{c}D_{a}(\nabla_kf)-\frac{h^{mn}}{2}\left(D_{a}v^{(k)}_{cn}+D_{c}v^{(k)}_{na}-D_{n}v^{(k)}_{ca}\right)(\partial_{m}H(\phi)),&&\notag\\
  &\nabla_k{\Delta H(\phi)}=\Delta(\nabla_kf)-h^{cd}(\partial_{c}
f)(\text{Div}v^{(k)})_{d}+h^{cd}(\partial_{c}f) D_{d}\theta_k-{v^{(k)}}^{ab}(D_{a}D_{b}f).&&
\end{align}

\item The Raychaudhuri equation for the shear is
\begin{equation}
\mathcal{R}_{kcka}=-\frac{D\sigma^k_{ac}}{D\lambda}+\frac{1}{4}v^{(k)}_{ab}{v^{(k)}}^b_{c}-\frac{h_{ac}}{(D-2)}\frac{d\theta_k}{d\lambda}-\frac{\theta_k}{(D-2)}v^{(k)}_{ac}.\label{sigray}
\end{equation}

\end{enumerate}

\section{Black brane solution in Sakai-Sugimoto model}\label{sakai}
We consider the 5-dimensional black brane solution in the Sakai-Sugimoto model
\begin{equation}
 ds^2=-r^{\frac{5}{3}}\Delta(r)dt^2+\frac{dr^2}{r^{\frac{4}{3}}\Delta(r)}+r^{\frac{5}{3}}(dx^idx_i).
\end{equation}
Here $\Delta(r)=1-\left(\frac{r_h}{r}\right)^3$. Now, using the following set of transformations, one can rewrite this metric in the Kruskal-Szekeres (null) coordinates by defining the $(U, V)$ coordinates. 
% \begin{equation}
%  du=dt-e^{w(r)/2}\frac{dr}{h(r)},~~dv=dt+e^{w(r)/2}\frac{dr}{h(r)}
% \end{equation}
% Therefore
% \begin{flalign}
%  &dudv=\left(dt-e^{w(r)/2}\frac{dr}{h(r)}\right)\left(dt+e^{w(r)/2}\frac{dr}{h(r)}\right)=dt^2-e^{w(r)}\frac{dr^2}{h(r)^2}&&\notag\\
%   &\Rightarrow -h(r)e^{-w(r)}dudv=-h(r)e^{-w(r)}dt^2+\frac{dr^2}{h(r)}&&
% \end{flalign}
\begin{align}
& U=-e^{-\kappa u},~~V=e^{\kappa v},~u=t-\int\frac{dr}{r^{\frac{3}{2}}\Delta(r)},~v=t+\int\frac{dr}{r^{\frac{3}{2}}\Delta(r)}.&&\notag
% & dU=\kappa e^{-\kappa u}du=-\kappa Udu,~~dV=\kappa e^{\kappa v}dv=\kappa Vdv\Rightarrow dUdV=-\kappa^2 UVdudv&&\notag\\
% &\Rightarrow \frac{h(r(U,V))e^{-w(r(U,V))}}{\kappa^2 UV}dUdV=-h(r(u,v))e^{-w(r(u,v))}dudv=-h(r)e^{-w(r)}dt^2+\frac{dr^2}{h(r)}&&\notag
\end{align}
This coordinate system is necessary for this analysis because in these coordinates, $V$ acts as an affine parameter along the null generator. The metric becomes
\begin{align}
ds^2=r^{\frac{5}{3}}\left(\frac{\Delta(r)}{\kappa^2 UV}dUdV+dx^2+dy^2+dz^2\right).\notag
\end{align}
Here $r=r(U,V)$.
% \begin{equation}
%  ln(-UV)=2\kappa\int e^{w(r)/2}\frac{dr}{h(r)}
% \end{equation}
%     Differentiating and equating coefficient of $dU$ and $dV$ from both sides gives
% \begin{equation}
%  \frac{\partial r}{\partial U}=e^{-w(r)/2}\frac{h(r(U,V))}{2\kappa U},~~ \frac{\partial r}{\partial V}=e^{-w(r)/2}\frac{h(r(U,V))}{2\kappa V}
% \label{differ1}
%  \end{equation}
The reason for shifting to the Kruskal-Szekeres coordinates is that the null vector gets identified with the coordinate basis vectors, and the coordinate behaves as an affine parameter. The null generators are $k=\partial_V$ and $n=-\frac{1}{g_{UV}}\partial_U$ on this background. 

The null expansion $\theta_k$ on the event horizon corresponding to the auxiliary null vector is given by
\begin{align}\label{nthet}
% & \theta_n=-\frac{6 U V\kappa^2 e^{w(U,V)}}{h(U,V)r(U,V)}\frac{\partial r}{\partial U}=\frac{3}{r(U,V)}\left(-\frac{2 U V\kappa^2 e^{w(U,V)}}{h(U,V)}\frac{\partial r}{\partial U}\right)&&\notag\\
% &=-\frac{6 U V\kappa^2 e^{w(r)}}{h(r)r_H}\left(e^{-w(r)/2}\frac{h(r)}{2\kappa U}\right)=-\frac{3\kappa V e^{w(r)/2}}{r_H}&&\notag\\
& \theta_k=\frac{5}{2r_h}\frac{\partial r}{\partial V}.
\end{align}
The rest of the null expansion and shear vanish on the stationary event horizon.

Further, for any scalar function, we have
\begin{equation}
 \nabla_{k}\phi=\frac{2r_h\phi'(r_h)}{5}\theta_k.
\label{scalfor}
\end{equation}

\section{Spherically symmetric static black hole metric in $D=5$}\label{ansatz}
The general spherically symmetric static black hole metric in $D=5$ is written as
\begin{equation}
 ds^2=-e^{-w(r)}f(r)dt^2+\frac{dr^2}{f(r)}+r^2(dx^2+dy^2+dz^2).
\end{equation}
For this metric, the surface gravity for the stationary event horizon is given by
\begin{equation}
 \kappa^2_h=-\frac{1}{2}(\nabla^a\xi^b)(\nabla_a\xi_b)\mid_{r=r_h}\Rightarrow \kappa_h=\frac{1}{2}f'(r)e^{-\frac{w(r)}{2}}.\label{surfg}
 \end{equation}
Here $\frac{\partial}{\partial_t}$ is the timelike Killing vector. 

The Kruskal-Szekeres coordinates metric is given by defining the $(U, V)$ coordinates
% \begin{equation}
%  du=dt-e^{w(r)/2}\frac{dr}{h(r)},~~dv=dt+e^{w(r)/2}\frac{dr}{h(r)}
% \end{equation}
% Therefore
% \begin{flalign}
%  &dudv=\left(dt-e^{w(r)/2}\frac{dr}{h(r)}\right)\left(dt+e^{w(r)/2}\frac{dr}{h(r)}\right)=dt^2-e^{w(r)}\frac{dr^2}{h(r)^2}&&\notag\\
%   &\Rightarrow -h(r)e^{-w(r)}dudv=-h(r)e^{-w(r)}dt^2+\frac{dr^2}{h(r)}&&
% \end{flalign}
\begin{align}
U=-e^{-\kappa u},~~V=e^{\kappa v}.\notag
% & dU=\kappa e^{-\kappa u}du=-\kappa Udu,~~dV=\kappa e^{\kappa v}dv=\kappa Vdv\Rightarrow dUdV=-\kappa^2 UVdudv&&\notag\\
% &\Rightarrow \frac{h(r(U,V))e^{-w(r(U,V))}}{\kappa^2 UV}dUdV=-h(r(u,v))e^{-w(r(u,v))}dudv=-h(r)e^{-w(r)}dt^2+\frac{dr^2}{h(r)}&&\notag
\end{align}
Then the metric becomes
\begin{align}
ds^2=e^{-w(r(U,V))}\frac{f(r(U,V))}{\kappa^2 UV}dUdV+r(U,V)^2(dx^2+dy^2+dz^2),
\end{align}
where $U, V$ satisfy
\begin{equation}
 \ln(-UV)=2\kappa\int e^{w(r)/2}\frac{dr}{h(r)}.
\end{equation}
%     Differentiating and equating coefficient of $dU$ and $dV$ from both sides gives
% \begin{equation}
%  \frac{\partial r}{\partial U}=e^{-w(r)/2}\frac{h(r(U,V))}{2\kappa U},~~ \frac{\partial r}{\partial V}=e^{-w(r)/2}\frac{h(r(U,V))}{2\kappa V}
% \label{differ1}
%  \end{equation}
The null generators are $k=\partial_V$, $n=-\frac{1}{g_{UV}}\partial_U=-\frac{2\kappa^2 U Ve^w}{h(r(U,V))}\partial_U$, and the null expansion $\theta_n$ is given by
\begin{align}
% & \theta_n=-\frac{6 U V\kappa^2 e^{w(U,V)}}{h(U,V)r(U,V)}\frac{\partial r}{\partial U}=\frac{3}{r(U,V)}\left(-\frac{2 U V\kappa^2 e^{w(U,V)}}{h(U,V)}\frac{\partial r}{\partial U}\right)&&\notag\\
% &=-\frac{6 U V\kappa^2 e^{w(r)}}{h(r)r_H}\left(e^{-w(r)/2}\frac{h(r)}{2\kappa U}\right)=-\frac{3\kappa V e^{w(r)/2}}{r_H}&&\notag\\
\theta_n=-\frac{(D-2)}{r_h}\kappa V e^{w(r_h)/2}.\label{nthet}
\end{align}
For any scalar function, we note that for $i=k,n$
\begin{equation}
 \nabla_{i}H(\phi)=\frac{r_hH'(\phi_h)\phi'(r_h)}{(D-2)}\theta_i.
\label{scalfor}
 \end{equation}

 \section{Null Raychaudhuri Equation for the Einstein Scalar Gauss-Bonnet Gravity}\label{app:raychaudhuri}
The null Raychaudhuri equation for the affinely parametrized ($k^{\nu}\nabla_{\nu}k^{\mu}=0$) null generator $k=\frac{\partial}{\partial\lambda},~k_{\mu}k^{\mu}=0$ is given by \cite{lewandowski2005quasi}
\begin{equation}
 \frac{d\theta_k}{d\lambda}=-R_{\mu\nu}k^{\mu}k^{\nu}-\frac{\theta^2_k}{(D-2)}-{\sigma^{(k)}}^{ab}\sigma^{(k)}_{ab}.
 \label{rayc1}
\end{equation}
Here $\theta_k$ is the expansion scalar, $\sigma^{(k)}_{ab}$ is the shear tensor, 
 and $\mathcal{R}_{\mu\nu}$ is the spacetime Ricci tensor. The equation of motion is given by
 \begin{equation}
 -\mathcal{R}_{\mu\nu}k^{\mu}k^{\nu}=-\alpha \mathcal{Q}_{\mu\nu}k^{\mu}k^{\nu}+4\alpha \mathcal{P}_{\mu\nu}k^{\mu}k^{\nu}-8\pi G_N T_{\mu\nu}k^{\mu}k^{\nu}.
 \label{EOM}
\end{equation}
 Here
 \begin{equation}
\mathcal{Q}_{\mu\nu}=-2\left(\mathcal{R}\mathcal{R}_{\omega\theta}-2\mathcal{R}_{\omega\mu}\mathcal{R}^{\mu}_{\theta}-2\mathcal{R}^{\mu\nu}\mathcal{R}_{\omega\mu\theta\nu}+\mathcal{R}_{\omega}^{\mu\nu\rho}\mathcal{R}_{\theta\mu\nu\rho}\right)+\frac{1}{2}g_{\mu\nu}+\frac{1}{2}g_{\mu\nu}\mathcal{R}^2_{GB},
\end{equation}
 and 
\begin{flalign}
 \mathcal{P}_{\mu\rho}=g^{\nu\alpha}g^{\sigma\beta}P_{\mu\nu\rho\sigma}(\nabla_{\alpha}\nabla_{\beta}H(\phi)).\notag
 \end{flalign}
 $P_{\mu\nu\rho\sigma}$ is the divergence-free $(\nabla_{\mu}P^{\mu}_{\nu\rho\sigma}=0)$ part of the Riemann tensor, given by
\begin{equation}
P_{\mu\nu\rho\sigma}=\left( \mathcal{R}_{\mu\nu\rho\sigma}-2g_{\mu[\rho}\mathcal{R}_{\sigma]\nu}+2g_{\nu[\rho}\mathcal{R}_{\sigma]\mu}+\frac{1}{2}g_{\mu[\rho}g_{\sigma]\nu}\mathcal{R}\right).
\label{interm}
\end{equation}
Now substituting \eqref{EOM} into \eqref{rayc1}, and decomposing the curvature terms using \eqref{metans}, then using the Gauss equation given in \eqref{gauss1}, we obtain the generalized null Raychaudhuri equation for Einstein-Scalar-Gauss-Bonnet gravity
 \begin{align}
&  \left(1+2\alpha H(\phi)\bar{R}-4\alpha\Delta^{(h)}H(\phi)\right)\nabla_k\theta_k=-\left(1+2\alpha H(\phi)\bar{R}-4\alpha\Delta^{(h)}H(\phi)\right)\frac{1}{4}{v^{(k)}}^{ab}{v^{(k)}}_{ab}&&\notag\\
&+\alpha\mathcal{R}_{\mu\nu}k^{\mu}k^{\nu}\Bigl(H(\phi)J+2\text{Tr}(W)\Bigr)-4\alpha H(\phi)\mathcal{R}_{kcka}\Bigl(\bar{R}^{ac}+\frac{1}{2}U^{ac}-\frac{1}{4}V^{ac}\Bigr)&&\notag\\
&+4\alpha \mathcal{R}_{kbka}\left(D^{b}D^{a}H(\phi)-\frac{1}{2}W^{ba}\right)+8\alpha h^{ab}\left(k^{\mu}D_{b}\nabla_{\mu} H(\phi)-\frac{h^{cd}}{2}v^{(k)}_{bc}D_dH(\phi)\right)h^{ef}\mathcal{R}_{kfae}&&\notag\\
&+2\alpha H(\phi)\mathcal{R}_{kdbn}\mathcal{R}_{k}^{~dbn}-4\alpha H(\phi)h^{cd}h^{ab}h^{mn}\mathcal{R}_{kacb}\mathcal{R}_{kmdn}&&\notag\\
&-2\alpha\biggl(\bar{R}+\frac{1}{2}J\biggr)k^{\mu}k^{\nu}(\nabla_{\mu}\nabla_{\nu}H(\phi))-8\pi G_N T_{\mu\nu}k^{\mu}k^{\nu}.&&
\label{raychan}
\end{align}
Here 
\begin{align}
&J=\left(4\theta_n\theta_k-{v^{(k)}}^{pq}v^{(n)}_{pq}\right),&&\notag\\
&W_{ab}=\left(v^{(n)}_{ab}k^{\mu}\nabla_{\mu}H(\phi)+v^{(k)}_{ab}n^{\mu}\nabla_{\mu}H(\phi)\right),&&\notag\\
&U_{bd}=\left(v^{(k)}_{bd}\theta_n+v^{(n)}_{bd}\theta_k\right),&&\notag\\
&V_{bd}=\left({v^{(k)}}_{d}^{p}v^{(n)}_{pb}+{v^{(n)}}_{d}^{p}v^{(k)}_{pb}\right).
 \end{align}
The expressions for $U_{bd}$ and $V_{bd}$ in terms of expansion are given below
\begin{align}
&U_{bd}=\left(v^{(k)}_{bd}\theta_n+v^{(n)}_{bd}\theta_k\right)=2\left(\sigma^{(k)}_{bd}\theta_n+\sigma^{(n)}_{bd}\theta_k+\frac{2\theta_n\theta_k h_{bd}}{(D-2)}\right),&&\notag\\
&V_{bd}=\left({v^{(k)}}_{d}^{p}v^{(n)}_{pb}+{v^{(n)}}_{d}^{p}v^{(k)}_{pb}\right)=4\left({\sigma^{(k)}}_{d}^{p}\sigma^{(n)}_{pb}+{\sigma^{(n)}}_{d}^{p}\sigma^{(k)}_{pb}+2\frac{\sigma^{(n)}_{db}\theta_k}{(D-2)}+2\frac{{\sigma^{(k)}}_{bd}\theta_n}{(D-2)}+2\frac{h_{bd}\theta_k\theta_n}{(D-2)^2}\right).&&\notag
\end{align}
Now we define $J$ as follows
\begin{align}
 &J=\frac{1}{2}\left(2\text{Tr}(U)-\text{Tr}(V)\right)=4\theta_n\theta_k-{v^{(k)}}^{ab}v^{(n)}_{ab}=-4{\sigma^{(n)}}^{ab}\sigma^{(k)}_{ab}+4\frac{(D-3)}{(D-2)}\theta_k\theta_n.&&\notag
\end{align}
% \begin{equation}
%  k^{\mu}\nabla_{\mu}H(\phi)=\frac{r_hH'(\phi_h)\phi'(r_h)}{(D-2)}\theta_k=\frac{c}{(D-2)}\theta_k,~ n^{\nu}\nabla_{\nu}H(\phi)=\frac{r_hH'(\phi_h)\phi'(r_h)}{(D-2)}\theta_n=\frac{c}{(D-2)}\theta_n
% \end{equation}
Using \eqref{scalfor}, we see that $W_{ab}$ is related to $U_{ab}$ as follows
\begin{align}
& W_{ab}=\left(v^{(n)}_{ab}k^{\mu}\nabla_{\mu}H(\phi)+v^{(k)}_{ab}n^{\mu}\nabla_{\mu}H(\phi)\right)&&\notag\\
&=\frac{c}{(D-2)}\left(v^{(n)}_{ab}\theta_k+v^{(k)}_{ab}\theta_n\right)=\frac{c}{(D-2)}U_{bd}&&\notag\\
&\Rightarrow \text{Tr}(W)=\frac{c}{(D-2)}\text{Tr}(U)=\frac{c}{(D-2)}\left(2\theta_k\theta_n+2\theta_n\theta_k\right)=\frac{4c}{(D-2)}\theta_k\theta_n.&&\notag
\end{align}

\section{The second variation}\label{thetakpDk}
The covariant derivative of $\tilde{K}$ is given by
\begin{align}
 &\nabla_k\tilde{K}=(\nabla_k\theta_k)(1+2\alpha H(\phi)\bar{R}-4\alpha\Delta^{(h)} H(\phi))+2\alpha \left(H(\phi) \theta_k+(\nabla_kH(\phi))\right)(\nabla_k\bar{R})\cr
 &-2\alpha H(\phi){v^{(k)}}^{e b}(\nabla_k\bar{R}_{e b})-4\alpha\theta_k\nabla_k(\Delta^{(h)} H(\phi))+2\alpha\nabla_k(D_{a}D_{c}H(\phi)){v^{(k)}}^{ac}\cr
 &+2\alpha\left(\bar{R} \theta_k-\bar{R}_{e b}{v^{(k)}}^{e b}\right)(\nabla_k H(\phi))+2\alpha(\nabla_k\nabla_kH(\phi))\bar{R}-3\alpha(D_{e}D_{b}H(\phi))h^{bn}{v^{(k)}}^{em}{v^{(k)}}_{m n}\cr
 &-4\alpha(D_{e}D_{b}H(\phi))h^{em}h^{bn}\mathcal{R}_{knkm}
 +3\alpha H(\phi)h^{bn}\bar{R}_{eb}{v^{(k)}}^{em}{v^{(k)}}_{m n}+4\alpha H(\phi)h^{em}h^{bn}\bar{R}_{eb}\mathcal{R}_{knkm}.&&\notag
\end{align}
Now, the first term in the above expression can be replaced by the generalized Raychaudhuri equation given by \eqref{raychan}. Substituting this, we get
\begin{flalign}
 &\nabla_k\tilde{K}=-(1+2\alpha H(\phi)\bar{R}-4\alpha\Delta^{(h)}H(\phi))\frac{1}{4}{v^{(k)}}^{ab}{v^{(k)}}_{ab}+\alpha\mathcal{R}_{\mu\nu}k^{\mu}k^{\nu}\Bigl( H(\phi)J+2\text{Tr}(W)\Bigr)&&\notag\\
 &-\alpha H(\phi) h^{ab}h^{cd}\mathcal{R}_{kcka}\Bigl(2U_{bd}-V_{pd}\Bigr)-2\alpha h^{fd}h^{ec}\mathcal{R}_{kekf} W_{cd}-\alpha J(\nabla_{k}\nabla_{k}H(\phi))&&\notag\\
&+3\alpha H(\phi)h^{bn}\bar{R}_{eb}{v^{(k)}}^{em}{v^{(k)}}_{m n}+2\alpha\left(\bar{R} \theta_k-\bar{R}_{e b}{v^{(k)}}^{e b}\right)(\nabla_k H(\phi))-8\pi G_N T(k,k)&&\notag\\
&+{2\alpha H(\phi)h^{ab}h^{mn}h^{cd}\mathcal{R}_{kdbn}\mathcal{R}_{kcam}}-{4\alpha H(\phi)h^{cd}h^{ab}h^{mn}\mathcal{R}_{kacb}\mathcal{R}_{kmdn}-2\alpha H(\phi){v^{(k)}}^{e b}(\nabla_k\bar{R}_{e b})}&&\notag\\
&+{8\alpha h^{ab}h^{cd}\mathcal{R}_{kacb}\left(D_{d}\nabla_k H(\phi)-\frac{h^{em}}{2}v^{(k)}_{de}D_mH(\phi)\right)}+{2\alpha \biggl(H(\phi) \theta_k+(\nabla_kH(\phi))\biggr)(\nabla_k\bar{R})}&&\notag\\
 &-{4\alpha\theta_k\nabla_k(\Delta^{(h)} H(\phi))+2\alpha\nabla_k(D_{a}D_{c}H(\phi)){v^{(k)}}^{ac}}-{3\alpha(D_{e}D_{b}H(\phi))h^{bn}{v^{(k)}}^{em}{v^{(k)}}_{m n}}.&&
 \end{flalign}
The curvature terms can be decomposed using the Codazzi equation and using other identities in Appendix \ref{ident}. We can simplify these terms to obtain the following expression
\begin{align}
 &\theta_k\tilde{K}+\nabla_k\tilde{K}=-8\pi G_N T_{\mu\nu}k^{\mu}k^{\nu}+\alpha\mathcal{R}_{\mu\nu}k^{\mu}k^{\nu}\Bigl(H(\phi)J+2\text{Tr}(W)\Bigr)\cr
 &-\alpha H(\phi) h^{ab}h^{cd}\mathcal{R}_{kcka}\Bigl(2U_{bd}-V_{bd}\Bigr)-2\alpha \mathcal{R}_{kekf}W^{ef}-\alpha (\nabla_{k}\nabla_{k}H(\phi))J
 \cr
 &+{\alpha\left(4\theta_k{v^{(k)}}^{eb}-2h^{bd}{v^{(k)}}^{ec}{v^{(k)}}_{c d}\right)(D_{e}D_{b}H(\phi))}\cr
 &{+(1+2\alpha H(\phi)\bar{R}-4\alpha\Delta^{(h)}H(\phi))\left(\theta^2_k-\frac{1}{4}{v^{(k)}}^{ab}{v^{(k)}}_{ab}\right)+2\alpha\bar{R} \theta_k(\nabla_k H(\phi))}\cr
 &{-4\alpha\bar{R}_{e b}{v^{(k)}}^{e b}(\nabla_k H(\phi))}{+2\alpha H(\phi)h^{bn}\bar{R}_{eb}{v^{(k)}}^{em}{v^{(k)}}_{m n}-\alpha H(\phi){v^{(k)}}^{e b}\bar{R}_{r e bc}{v^{(k)}}^{rc}}\cr
 &{-2\alpha H(\phi)\theta_k{v^{(k)}}^{e b}\bar{R}_{e b}-2\alpha\theta_k H(\phi){v^{(k)}}^{e b}\bar{R}_{e b}}&&\notag\\
&{+2\alpha\theta_k(\nabla_kH(\phi))\bar{R}}+\text{Total Divergence}.&&\label{stilde1}
\end{align}
Finally, after computing the variation, we can impose conditions for a planar black hole in the above expression. The following two conditions hold: a) $\phi\equiv\phi(r)$ and b) $\bar{R}_{abcd}=0$. The first condition sets all the tangent derivatives of $H(\phi)$ to zero, and the second condition sets all curvature terms to zero. This simplifies the above expression drastically:
\begin{align}\label{stilde1}
 &\theta_k\tilde{K}+\nabla_k\tilde{K}=-8\pi G_N T_{\mu\nu}k^{\mu}k^{\nu}+\alpha\mathcal{R}_{\mu\nu}k^{\mu}k^{\nu}H(\phi)J-\alpha H(\phi) \mathcal{R}_{kcka}\Bigl(2U^{ac}-V^{ac}\Bigr)\cr
 &+2\alpha\left(\mathcal{R}_{\mu\nu}k^{\mu}k^{\nu}\text{Tr}(W)-\mathcal{R}_{kekf}W^{ef}\right)-\alpha k^{\mu}k^{\nu} (\nabla_{\mu}\nabla_{\nu}H(\phi))J+\left(\frac{(D-3)}{(D-2)}\theta^2_k-{\sigma^{(k)}}^{2}\right)\cr
 &+\text{Total Divergence}.
\end{align}
Before proceeding, we must first resolve the curvature terms containing the normal $k$. This is achieved by using \eqref{sigray}, \eqref{scalfor}, and \eqref{rayc1} as follows 
\begin{flalign}
&\mathcal{R}_{kcka}=-\frac{D\sigma^k_{ac}}{D\lambda}+\frac{1}{4}v^{(k)}_{ab}{v^{(k)}}^b_{c}-\frac{h_{ac}}{(D-2)}\frac{d\theta_k}{d\lambda}-\frac{\theta_k}{(D-2)}v^{(k)}_{ac},&&\notag\\
&\mathcal{R}_{\mu\nu}k^{\mu}k^{\nu}=-\frac{d\theta_k}{d\lambda}-\frac{1}{4}{v^{(k)}}^{ab}{v^{(k)}}_{ab}=-\left(\frac{d\theta_k}{d\lambda}+{\sigma^{(k)}}^{ab}{\sigma^{(k)}}_{ab}+\frac{\theta^2_k }{(D-2)}\right),&&\notag\\
&k^{\mu}k^{\nu}(\nabla_{\mu}\nabla_{\nu}H(\phi))=\frac{c}{(D-2)}\frac{d\theta_k}{d\lambda}.&&\notag
\end{flalign}
Before we proceed with our approximation, we note that all the geometric quantities $J,~W_{ab},~U_{ab},V_{ab}$ given in Appendix \ref{app:raychaudhuri} are already first order in $\theta_k,\sigma^{(k)}$, so only the derivatives of $\theta_k$ and $\sigma^{(k)}$ will contribute; the rest of the terms are higher than second order. Using this approximation, we get
\begin{align}
 &\theta_k\tilde{K}+\nabla_k\tilde{K}=-8\pi G_N T_{\mu\nu}k^{\mu}k^{\nu}+\alpha H(\phi)\Bigl(2U^{ac}-V^{ac}\Bigr)\frac{D\sigma^k_{ac}}{D\lambda}-\frac{(D-4)}{(D-2)}\alpha H(\phi)J\frac{d\theta_k}{d\lambda}&&\notag\\
 &+2\alpha W^{ac}\frac{D\sigma^k_{ac}}{D\lambda}-2\alpha\frac{(D-3)\text{Tr}(W)}{(D-2)}\frac{d\theta_k}{d\lambda}-\frac{\alpha J c}{(D-2)}\frac{d\theta_k}{d\lambda}+\left(\frac{(D-3)}{(D-2)}\theta^2_k-{\sigma^{(k)}}^{2}\right)&&\notag\\
 &+\text{Total Divergence}.&&\label{stilde1}
\end{align}
Furthermore, from Appendix \ref{app:raychaudhuri}, we observe that at second order in $\theta_k,\sigma^{(k)}$:
\begin{align}
%    &J=\left(4\theta_n\theta_k-{v^{(k)}}^{pq}v^{(n)}_{pq}\right),~~W_{ab}=\left(v^{(n)}_{ab}k^{\mu}\nabla_{\mu}H(\phi)+v^{(k)}_{ab}n^{\mu}\nabla_{\mu}H(\phi)\right)&&\notag\\
%    &U_{bd}=\left(v^{(k)}_{bd}\theta_n+v^{(n)}_{bd}\theta_k\right),~~V_{bd}=\left({v^{(k)}}_{d}^{p}v^{(n)}_{pb}+{v^{(n)}}_{d}^{p}v^{(k)}_{pb}\right)&&\notag\\
% &\text{Tr}(U)=4\theta_k\theta_n,~~\text{Tr}(V)=2{v^{(k)}}^{pb}v^{(n)}_{pb},~~2\text{Tr}(U)-\text{Tr}(V)=2(4\theta_k\theta_n-{v^{(k)}}^{ab}v^{(n)}_{ab})=2J&&\notag\\
% & \text{Tr}(W)=\frac{c}{(D-2)}\text{Tr}(U)=\frac{c}{(D-2)}\left(2\theta_k\theta_n+2\theta_n\theta_k\right)=\frac{4c}{(D-2)}\theta_k\theta_n&&\notag\\
&\Bigl(2U^{ac}-V^{ac}\Bigr)\frac{D\sigma^k_{ac}}{D\lambda}\simeq2\Bigl(2{\sigma^{(k)}}^{ac}\theta_n-4\frac{{\sigma^{(k)}}^{ac}\theta_n}{(D-2)}\Bigr)\frac{D\sigma^k_{ac}}{D\lambda}=4\frac{(D-4)}{(D-2)}\theta_n{\sigma^{(k)}}^{ac}\frac{D\sigma^k_{ac}}{D\lambda},&&\notag\\
&J\frac{d\theta_k}{d\lambda}=\left(4\theta_n\theta_k-{v^{(k)}}^{pq}v^{(n)}_{pq}\right)\frac{d\theta_k}{d\lambda}\simeq 4\frac{(D-3)}{(D-2)}\theta_k\theta_n\frac{d\theta_k}{d\lambda}.&&\notag
\end{align}
Using this approximation in the previous expression, we finally obtain
\begin{align}
 &\lambda\left(\theta_k\tilde{K}+\nabla_k\tilde{K}\right)=-8\pi G_N T_{\mu\nu}\lambda k^{\mu}k^{\nu}+\frac{d}{d\lambda}(\lambda\theta_nA)-A\theta_n-A\lambda\frac{d\theta_n}{d\lambda} &&\notag\\
 &+\frac{d}{d\lambda}(\lambda\theta_nB)-B\theta_n-B\lambda\frac{d\theta_n}{d\lambda}+\lambda\left(\frac{(D-3)}{(D-2)}\theta^2_k-{\sigma^{(k)}}^{2}\right)+\text{Total Divergence},&&\label{stilde112}
\end{align}
where we denote
\begin{equation}
 A=2\alpha\frac{(D-4)}{(D-2)}H(\phi_h)\left({\sigma^{(k)}}^{2}-\frac{(D-3)}{(D-2)}\theta^2_k\right),~B=\frac{2\alpha c}{(D-2)}\left({\sigma^{(k)}}^{2}-3\frac{(D-3)}{(D-2)}\theta^2_k\right).
\end{equation}
Now, from \eqref{nthet}, we know that
\begin{equation}
 \theta_n=-(D-2)\frac{\kappa_h \lambda e^{w(r_h)/2}}{r_h}.
\end{equation}
Furthermore, we note that 
\begin{equation}
-8\pi G_nT_{\mu\nu}\lambda k^{\mu}k^{\nu}=\frac{1}{2}(\nabla_k\phi)^2-8\pi G_n T^{\text{Gauge}}_{\mu\nu} k^{\mu}k^{\nu}.
\end{equation}
Using this in the previous expression, we get
\begin{align}
 &\lambda\left(\theta_k\tilde{K}+\nabla_k\tilde{K}\right)=4\alpha(D-4)\frac{\kappa_h \lambda e^{w(r_h)/2}}{r_h}H(\phi_h)\left({\sigma^{(k)}}^{2}-\frac{(D-3)}{(D-2)}\theta^2_k\right) &&\notag\\
 &+4\alpha c\frac{\kappa_h e^{w(r_h)/2}}{r_h}\left({\sigma^{(k)}}^{2}-3\frac{(D-3)}{(D-2)}\theta^2_k\right)\lambda+\lambda\left(\frac{(D-3)}{(D-2)}\theta^2_k-{\sigma^{(k)}}^{2}\right)&&\notag\\
 &-\frac{r^2_h}{2(D-2)^2}\phi'(r)^2\theta^2_k+\frac{d}{d\lambda}(\lambda\theta_nA)+\frac{d}{d\lambda}(\lambda\theta_nB)-8\pi G_N\lambda T_{\mu\nu} k^{\mu}k^{\nu}+\text{Total Divergence}.&&\notag\\
 \end{align}
 This is the expression we used in the main text.

%\end{appendices}

% Bibliography

%% [A] Recommended: using JHEP.bst file
%\bibliographystyle{JHEP}
%\bibliography{biblio.bib}

\providecommand{\href}[2]{#2}\begingroup\raggedright\endgroup

\end{document}